# Thriving Innovation Ecosystems: Synergy Among Stakeholders, Tools, and People

Shruti Misra and Denise Wilson

An innovation ecosystem is a multi-stakeholder environment, where different stakeholders interact to solve complex socio-technical challenges. We explored how stakeholders use digital tools, human resources, and their combination to gather information and make decisions in innovation ecosystems. To comprehensively understand stakeholders' motivations, information needs and practices, we conducted a three-part interview study across five stakeholder groups (N=13) using an interactive digital dashboard. We found that stakeholders were primarily motivated to participate in innovation ecosystems by the potential social impact of their contributions. We also found that stakeholders used digital tools to seek "high-level" information to scaffold initial decision-making efforts but ultimately relied on contextual information provided by human networks to enact final decisions. Therefore, people, not digital tools, appear to be the key source of information in these ecosystems. Guided by our findings, we explored how technology might nevertheless enhance stakeholders' decision-making efforts and enable robust and equitable innovation ecosystems.

## 1 INTRODUCTION

Technological innovation is key to fueling the U. S's economic growth and maintaining its global competitive edge [1]. In the U.S, innovative activities are concentrated in a few major metropolitan hubs, the key one being Silicon Valley. Globally, public, and private sector leaders have attempted to create similar hubs to fuel technological innovation in their region and mimic the economic prosperity that has accompanied the success of Silicon Valley [2,3]. However, innovation hubs comprise of complex, historical, and ever-evolving interactions between various stakeholders, which are difficult to replicate [4,5]. These interactions depend on many overlapping contexts such as the location where the interactions are occurring, the industrial sector that the stakeholders are embedded in and the technology itself. The diverse networks of interacting stakeholders whose primary goal is to further technological development and are often geographically collocated form an innovation ecosystem.

Innovation ecosystems are multi-stakeholder environments whose scale, emergent dynamics and path-dependent nature makes them challenging to study and model. Nonetheless, scholars have developed metrics and frameworks to characterize the rich stakeholder dynamics that underlie these ecosystems [6,7]. However, these theories and empirical metrics do not adequately capture how stakeholder dynamics unfold in practice. The gap between theory and practice exists primarily because (1) most existing metrics and theories were developed by researchers and economists for other researchers and economists [8,9,10]. They hardly address the needs and experiences of other stakeholders and practitioners in the ecosystem, (2) theories characterizing specific stakeholder interactions are scattered across multiple disciplines (innovation studies, economics, business management, entrepreneurship etc.), providing an non-cohesive narrative of how stakeholders operate in innovation ecosystems and, (3) there is a lack in understanding of how different stakeholders make actionable decisions by interacting with each other, using digital data and a host of digital tools which facilitate interactions between people and information. To that end, the goal of our study is to provide a cohesive narrative of stakeholder interactions and needs by conducting a multi-stakeholder analysis of motivations, perspectives, information needs and practices in an innovation ecosystem. We seek to answer the following research questions:

1. **RQ1:** How do different stakeholders seek to contribute to innovation ecosystems?
2. **RQ2:** What types of information and from which sources do different stakeholders seek to support these contributions?
3. **RQ3:** What are the design considerations for potential digital tools that can enhance stakeholder interactions, decision-making and contributions to an innovation ecosystem?

We prototyped an interactive data dashboard that visualized widely accepted metrics used to characterize activities in an innovation ecosystem. The dashboard tool was used for a three-part semi-structured interview study to elicit discussion with stakeholders and address the research questions. We interviewed 13 participants from six stakeholder groups derived from literature [11]: university stakeholders, corporate stakeholders, entrepreneurs, government stakeholder and "other". The interview had three-parts: Part 1 focused on stakeholders' perspectives, motivations, activities, and decision-making practices in their current role. Part 2 examined reactions to the dashboard tool through a think aloud activity. In the final part of the interview, participants reflected on the think aloud activity and offered suggestions on the usability and design of future iterations of the tool.

We used thematic analysis to identify five main themes related to: stakeholder motivations, interactions, activities, technology practices and, decision-making. Our findings highlight that most stakeholders are motivated to participate in an innovation ecosystem by the potential social impact of their contributions. This motivation grounds stakeholder interactions and activities. Further, stakeholders sought information along three main dimensions: information about people (key players in an ecosystem), information about money (how to raise and spend capital) and knowledge (scientific, business strategy etc.). A key finding of our study was that stakeholders leveraged digital tools to obtain "high-level" information that scaffolded further decision-making efforts. However, ultimately, stakeholders relied on the contextual information obtained from their personal, professional, and expert networks for final decision-making.

Our contributions lie at the intersection of innovation studies, human centered design, and visualization. They include (1) a framework for how stakeholders use digital and non-digital resources to operate in an innovation ecosystem and, (2) a set of design considerations that inform future dashboard tool development that supports decision-making of multiple stakeholder groups in an innovation ecosystem. To our knowledge, this is the first study that uses a multi-stakeholder approach to examine the synergy between digital and non-digital practices of stakeholders in an innovation ecosystem and suggests digital tool design recommendations to enhance them.

## 2 BACKGROUND

### 2.1 Innovation Ecosystems

"Innovation" is defined by the Organization for Economic Co-operation and Development (OECD) as "the implementation of a new or significantly improved product (good or service), or process, a new marketing method, or a new organizational method in business practices, workplace organization or external relations." [12]. Research and entrepreneurship are often considered to be key activities that are associated with innovation and are often localized in hubs such as Silicon Valley, Boston, Austin etc. These hubs comprise of various ways in which research, entrepreneurship and other innovative activities occur and evolve, thus forming an 'innovation ecosystem'. An innovation ecosystem "is the evolving set of actors, activities, and artifacts, and the institutions and relations, that are important for the innovative performance of an actor or a population of actors" [13]. These actors consist of material resources (knowledge, money, infrastructure etc.) and human resources (students, faculty, entrepreneurs etc.) who make up the institutions that participate in the ecosystem (universities, corporations, startups, VCs).

Successful innovation ecosystems are not messy groups of people, organizations, resources, and skills, but require proximity and cohesion between actors which is driven by collective action. Therefore, to understand how innovation ecosystems evolve and grow, it is necessary to understand the interactions that intertwine different actors (stakeholders) to form a coherent 'community' [14]. At the heart these 'communities' is the ability of diverse stakeholders to effectively communicate and interact across organizational and contextual boundaries. In this study, we focus on examining this flow of information and communication that occurs between stakeholders as mediated by digital and non-digital resources.

**2.2 Who are the key stakeholders in innovation ecosystems?**

In early works, three stakeholders were key to innovation ecosystems, namely universities, industry (corporations) and government. However, over the years as scholarship progressed, new stakeholders have emerged as being vital to innovation ecosystems. For instance, entrepreneurs are now considered as an important stakeholder because they are the ones developing and scaling products from scratch [15]. Other scholars have also observed that in practice, risk capital providers such as venture capitalists are also critical actors that lead growth and change in innovation ecosystems through funding [16]. We base our study on a five-stakeholder conceptual framework developed by researchers at Massachusetts Institute of Technology (MIT) based on historical analyses of well-developed ecosystems and global experiences with developing ecosystems [11]. We selected this conceptual model because it combines traditional stakeholders (such as universities, industry, and government) with newly recognized stakeholders such as entrepreneurs and venture capital to provide a more comprehensive categorization of all the actors in an innovation ecosystem. The five key stakeholders are:

1. **Entrepreneurs:** In the context of the model, entrepreneurs refer to individuals who form enterprises with the explicit intention to harness a competitive advantage based on new innovations (new technology, scientific insights, supply chains etc.) and quickly scale beyond local markets. Because of their experience on the "frontlines" of innovation, entrepreneurs motivate ecosystem-building efforts by communicating the support they need from other stakeholders. Therefore, entrepreneurs are critical actors in innovation ecosystems.
2. **Risk Capital:** This group refers primarily to investors who provide necessary funding to entrepreneurs to scale their business. Risk capital stakeholders not only fund promising startups but also provide mentorship and support. They are necessary to the ecosystem, as they accelerate the emergence of new innovative enterprises [16].
3. **Universities:** Universities play an important, multi-faceted role that varies widely across innovation ecosystems. They provide novel research, technical and entrepreneurship education, facilities, talented scientists, and students etc. A variety of actors within a university can participate in an innovation ecosystem from Technology Licensing/Transfer Office to individual faculty and students. The amount a university contributes to an innovation ecosystem varies. For example, MIT and Stanford have played a seminal role in establishing the Greater Boston and Silicon Valley innovation ecosystems respectively. However, that is not the case for many other prominent and the regions they belong to. Therefore, a strong university does not imply a strong innovation ecosystem [17].
4. **Corporations:** Large corporations play an important role in ecosystem-building by developing and attracting talent, contributing to risk capital through corporate venture funds and providing facilities such as laboratories and equipment that support the innovation infrastructure. For example, Google, Microsoft, and Amazon all offer special cloud credit programs that support startups and entrepreneurs and thus provide them with a platform to build on.

5. **Government:** The role of the government is critical in building innovation ecosystems. While government stakeholders may not be leaders, their role is to "set the table" and create rules and policy that enable robust ecosystems. For example, Mazzucato (2015) argues that the establishment of various government research programs and organizations (such as the Defense Advanced Research Projects Agency) during and post-World War II has fueled the development of most information technologies (including the computer) today [18].

In addition to the five key stakeholders discussed above, there are many *other* key players that play in an innovation ecosystem. These include accelerators and incubators that help startups grow by providing the training, mentorship and resources needed by entrepreneurs to scale their business. Examples also include lawyers and consultants who help various stakeholders negotiate intellectual property rights, conduct market or user research, build partnerships etc.

### 2.3 Tools and Technologies in Innovation Ecosystems

The rise of open and proprietary digital data presents an opportunity to understand the complex dynamics that underlie stakeholder activity in innovation ecosystems. Previously unavailable social and organizational data is now available through networking websites such as LinkedIn, press releases and other online databases. Furthermore, public data on patents, research and business activity is now easily accessible through online government databases. However, an innovation ecosystem is not characterized by a single source of data, but relies on multiple disjointed, misaligned and heterogenous sources of data, which are challenging to combine in meaningful ways [19].

Existing tools for ecosystem analysis include business intelligence tools [20,21], open-source visualization tools developed by various government organizations and scholars [22,23,24] and proprietary databases (such as Pitchbook and Crunchbase). While these tools provide a starting point, they only focus on one or two aspects of the innovation ecosystem. For example, the National Science Foundation's S&E indicator visualizations provide a suite of individual state-levels maps for various research and business indicators. However, there is no way to connect those indicators together into a cohesive ecosystem analysis. Similarly, maps focused on entrepreneurial and business activity [23,24] rarely consider research activity, ignoring a vital part of the ecosystem. Basole et al (2018) mention that existing ecosystem analysis tools do not reflect the interconnected nature of innovation ecosystems [19]. Similarly, Smith highlights a need for web mapping tools for socio-economic data that can simplify research and allow comparisons to be made between various innovation indicators in different locations [25].

Therefore, better tools are needed that integrate different aspects of the innovation ecosystem in an intuitive and user-friendly way and can enhance the decision-making activities of stakeholders. To develop such tools, it is vital to take a user-centered design approach because different stakeholders may use different data sources and tools based on their information needs, tasks, and comfort with technology (for instance, technically savvy analysts vs. non-technical program managers). Any tool being developed to address the information needs of stakeholders needs to cater to the diverse stakeholder profiles discussed in Section 2.2. To do so, it is important to understand how stakeholders operate in innovation ecosystems, how they interact with each other, their activities, information needs and practices. The confluence of (1) stakeholder diversity, (2) ubiquity (and disjointedness) of digital data that characterizes innovation ecosystems, and (3) lack of user-centered tools that address different stakeholder profiles while combining data in functional ways form the basis of the research questions addressed by this study.

# 3 METHODS

## 3.1 Participants

Table 1: Participant Roles and Stakeholder groups

| Participant # | Group[a] | Role | Notes |
|---|---|---|---|
| 1 | O & U | Accelerator Founder, Faculty | Also has experience in investing |
| 2 | C | Financial Business Partner | |
| 3 | C | Analyst | |
| 4 | O & C | Accelerator Associate Director, Sales Manager | |
| 5 | U | Program Director | University College of Engineering |
| 6 | ED | Program Manager | Economic Development Non-Profit |
| 7 | E | Startup Founder | |
| 8 | U | Innovation Manager | University Technology Transfer |
| 9 | U | Assistant Director | University Business School |
| 10 | ED & E | Startup Founder, Consultant | |
| 11 | E | Startup Founder | |
| 12 | E & R | Startup Founder, Senior Venture Partner | |
| 13 | U | Senior Technology Manager | University Technology Transfer |

[a]C=corporate stakeholders, E= entrepreneurs, ED= economic development, O=other (accelerator stakeholders), R=risk capital stakeholders, U=university stakeholders

A total of 13 participants were recruited for the study across six stakeholder groups working in Knowledge-Intensive Industries (KTI) as defined by the National Science Foundation [23], which primarily include science, technology, and healthcare sectors. These stakeholders included: 4 entrepreneurs, 5 university stakeholders, 2 government (economic development) stakeholders, 1 risk capital (investors) stakeholder, 3 corporate (analysts) stakeholders and 2 other stakeholders, specifically who manage accelerator programs. Entrepreneurs were individuals who founded early-stage (0-5 years old) technology companies. University stakeholders included individuals who are responsible for technical and business education programs and support university technology transfer activities. Government stakeholders were individuals who work with government organizations to support activities that are a direct result of organized economic development efforts. Risk capital stakeholders included individuals who have either invested capital in technology companies individually or have been partners in venture capital firms. Corporate stakeholders included participants who worked for private corporations as analysts or managers on expanding a corporation's customer or partner portfolio. We also interviewed two participants who manage an accelerator program that supports startups as they scale by providing them with access to mentors. They fall in the "Other" category of the stakeholder framework [11]. All participants were based in the Seattle area to prevent differences in cities as being a confounding factor. All interviews were conducted remotely on Zoom. Table 1 provides an overview of all the participants, their roles, and affiliations.

## 3.2 Interview Study

We conducted a three-part interview study to address our primary research questions. The interviews were conducted with a dashboard tool created in Tableau (Section 4). Participants were given a website link where the tool was hosted and asked to share their screen as they navigated through it. Each interview session lasted for around 1 hour during which the audio, video and screens of the participants were recorded. Appropriate IRB approval was obtained prior to recruiting for the interview. Each interview consisted of three parts:

1. **Part 1: Stakeholder Perspectives on Innovation Ecosystems:** The goal of this part was to understand how each participant experienced their region's innovation ecosystem and their responsibilities, activities, and practices within it. The participant was prompted to describe the local ecosystem and why they participated in it. Additional questions explored what types of information they used to augment their knowledge of the ecosystem and support the goals of their job. The participant was also prompted to discuss how they sought such information to effectively do their jobs. Part 1 also laid the foundation for exploring the digital dashboard tool designed to support stakeholders in innovation ecosystems during the next part of the interview.
2. **Part 2: Think Aloud Activity with the Data Visualization Probe:** This part of the interview sought to understand (a) what information within the dashboard tool were of most interest to the participant; and (b) how the participant would realistically navigate through the tool. The participant was given a general description of the dashboard and the data within it. The participant was then encouraged to explore the tool for as long as they wanted and were asked to verbalize their thoughts (and questions) as they explored. Particular attention was paid to the order in which participants visited the tabs, the duration of time they spent on each tab, and whether they visited all the tabs.
3. **Part 3: Reflections:** To wrap up the session, the participants reflected on their experience of using the data visualization tool. This part of the interview sought to understand (a) what additional information was of interest to the participant; (b) how existing information could be better presented to serve the needs of the participant; and (c) overall, how valuable the tool was to the participant in gathering information and supporting their contributions to innovation ecosystems.

### 3.3 Data Analysis

The audio and screenshare recordings were transcribed and thematically analyzed. The first round of coding was done in an inductive manner to discover themes and patterns that were common among stakeholders in each group. The next round of coding was based on the constant comparative approach, where themes emerging from the first round were compared across different stakeholder groups to create categories, establish, and refine boundaries of those categories, and summarize the content of each category [26]. Five key themes emerged from the coding process: stakeholder motivations, interactions, information activities, tools and resources used and information needs (Table 2).

Table 2: Summary of key themes that emerged

| # | Theme | Description |
|---|---|---|
| 1 | Motivation | Why do stakeholders participate in an innovation ecosystem? |
| 2 | Stakeholder Interactions | What is the relationship between different stakeholders in an innovation ecosystem? |
| 3 | Information Activities | How to stakeholders participate and make-decisions in an innovation ecosystem? |
| 4 | Tools and Resources | What digital and non-digital tools do stakeholders use to participate and make decisions in an innovation ecosystem? |
| 5 | Information Needs | What information and data do stakeholders rely on to make participate and make decisions in an innovation ecosystem? |

## 4 DIGITAL DASHBOARD TOOL DESIGN

The data dashboard tool was developed with the goal of having an interactive medium for stakeholders to explore existing quantitative metrics used to characterize innovation ecosystems and examine how relevant they are to diverse stakeholder groups. The metrics were selected based on literature that specified how innovation activities are measured in an ecosystem [7, 11, 27]. The tool was prototyped as a Tableau Dashboard, where the main selection criteria was city-level (also known as metropolitan statistical area level (MSA)). We selected MSA as the unit of analysis because innovation ecosystems are usually centered in urban, metropolitan regions. The tool provides a snapshot of different metrics for the year 2019. To control scope, we only visualized cross-sectional data. Table 3 summarizes the data types and sources visualized for each MSA for the year 2019.

Table 3: Data types, sources, and description for the dashboard tool

| Data Type (unit) | Source | Activity Type |
|---|---|---|
| Federal R&D Funding ($) | Grants.gov | Indicator for research activity |
| Small Business Innovation (SBIR)/Small Business Technology Transfer (STTR) grant funding ($) | Grants.gov | Indicator for research commercialization from universities to industry |
| Total venture capital funding ($) | Pitchbook | Indicator for entrepreneurship |
| Number of new investors who made deals | Pitchbook | Indicator for entrepreneurship |
| Number of new firms that emerged by industry sector | Business Census Data (Bureau of Labor Statistics) | Indicator for economic development |
| Venture Capital Funding by sector (%) | Pitchbook | Indicator for entrepreneurship |
| Venture Capital Funding by technology (%) | Pitchbook | Indicator for entrepreneurship |
| Employment created by industry sector | Bureau of Labor Statistics | Indicator for economic development |
| University research grant funding by institution and discipline | National Science Foundation | Indicator for research activity |

## 4.1 Tool Design

Our tool design was informed by basic visualization literature [28,29]. The goal was to follow the visualization mantra as proposed by Shneiderman (1996), which is: "overview first, zoom and filter, then details-on-demand" [28]. Further, the goal of this first iteration of tool design was to understand the different mode of analyses adopted by different stakeholders. Therefore, a combination of pie charts, heatmaps, stacked bar charts and map-based visualization techniques were used. We wanted to understand whether stakeholders preferred a "top-down" analytical approach (expressed by stacked bar charts), a "bottom-up" one (expressed by pie charts) or a combination thereof. A map-based visualization was used to assess whether stakeholders cared about "seeing" the geographical context of innovation activities. Further, heatmap representations were used to compare different types of funding across MSAs to gauge whether stakeholders were interested in ranking different innovation ecosystems. Finally, we created 8 visualizations across 4 chart types (Figure 1).

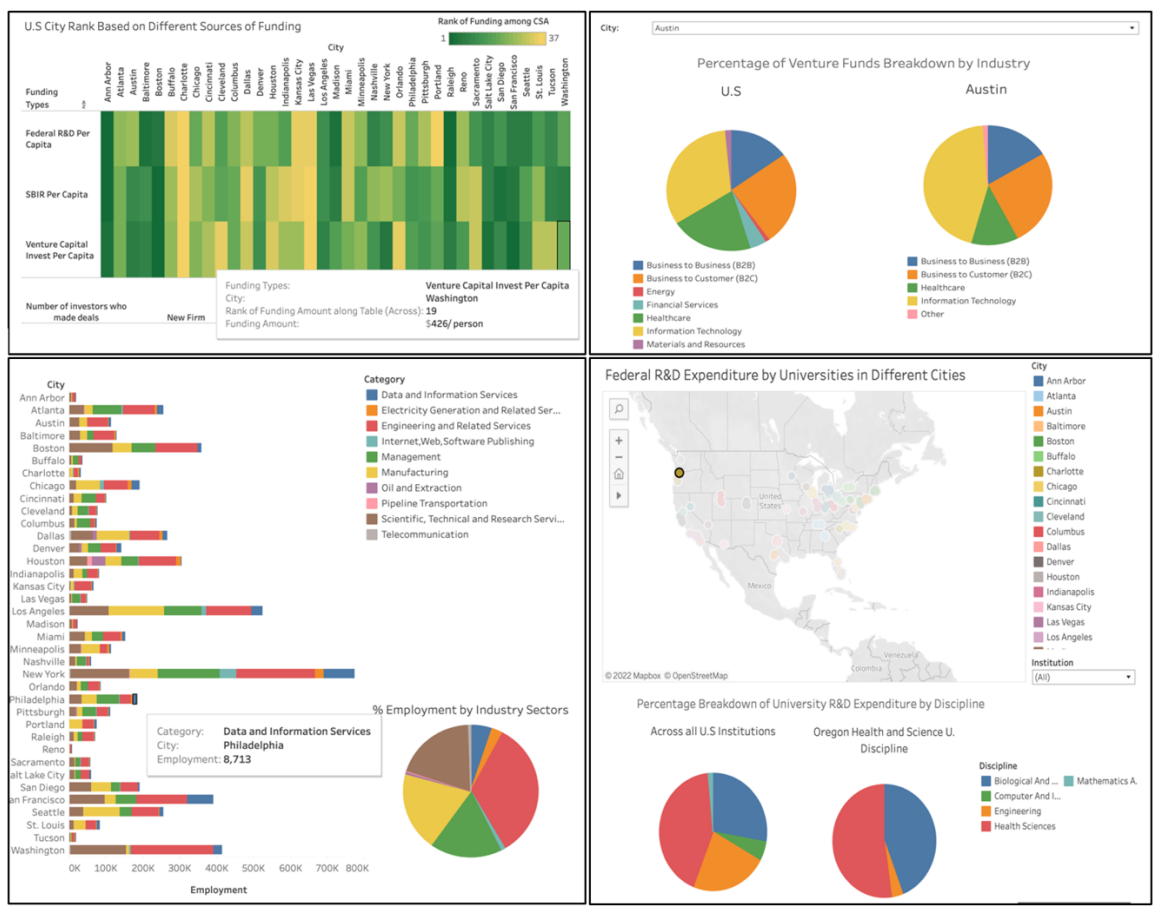

Figure 1: Different types of charts used in the dashboard tool

## 5 RESULTS

### 5.1 Stakeholder Perspectives

*5.1.1 Motivation*

To understand the underlying lens with which different stakeholders view an innovation ecosystem and how they sought to contribute to it, motivations for their contributions were examined. We found that 12 of 13 participants (92.3%) were motivated by the social impact of their contributions to innovation and innovation ecosystems. For example, one university stakeholder stated:

> "My university technology transfer job felt more meaningful [as compared to a previous for-profit job] because I was looking at research and was helping people develop new products that have societal value."

Three sub-categories of social impact emerged as being prominent motivators: technological advancement, equity, and economic development.

*Technological Advancement (N=6):* The most common sub-category through which participants wanted to generate social impact was to develop technologies that solve pressing societal needs. For example, one corporate participant decided to work with healthcare technologies because one of his children had delayed developmental milestones. The direct impact that healthcare technologies have on human life motivated him to manage the healthcare business portfolio for a large corporation. Similarly, another participant was interested in technology commercialization of university research technologies because she was driven by the question:

"How do you create, grow and scale massively scalable technology that has societal impact?"

*Equity (N=3):* Some interview participants wanted to create social impact by enabling equitable outcomes. There were two distinct angles to how equity motivated participants. Some participants were interested in creating technologies that addressed inequities in social systems. For example, one participant co-founded a company that sells a more efficient version of an existing technology in the hopes of lowering system-wide costs and financial barriers to accessing healthcare for low-income communities. Other participants were driven by the need to support underrepresented entrepreneurs, researchers, and student-entrepreneurs in their activities. For example, one participant mentioned that:

> "A huge chunk of the [participant's university] students are first generation college students. So, a dedicated investment space that looks at the university's alum means that there's a dedicated space looking at people who are likely to be first generation alum who might not have access to those friends and family capital networks for that early stage- money".

By creating the above investment space through her venture capital organization, the participant wanted to level the playing field for those who do not have the generational and social capital to pursue innovative activities and participate in the region's innovation ecosystem.

*Economic Development (N=2):* Another key area social impact that motivated some participants was economic development. These participants were undertaking activities that they thought directly contributed to the economic prosperity of the region in the form of job creation, GDP growth etc. For example, one participant commented on his motivation for bringing investment and new companies to the region of interest:

> "My primary interest was the Japanese economic modernization experience, how unique that was and what modern economies can learn from the Japanese experience. The above led me to government and international trade and investment work."

Only one participant did not appear motivated by social impact. This participant was motivated by wanting to gain skills and experience by working in economic development.

*5.1.2 Interactions*

The primary motivation frames stakeholder interactions in an innovation ecosystem. For example, when a university stakeholder interacts with corporate stakeholders for research partnerships, the primary motivation is technological advancement. However, when the same stakeholders interact for educational purposes, the primary motivation might be around economic development or equitable talent development. While the same stakeholder groups are interacting in both activities, the individual actors involved in the interactions likely belong to very different departments in each stakeholder organization. Figure 2 summarizes the interactions between the different stakeholders that emerged from interviews. This framework of interactions adds to prior frameworks described in literature [11,6].

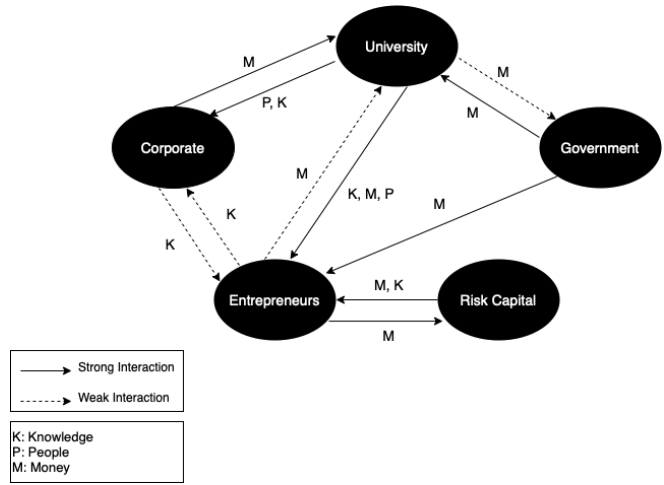

Figure 2: Stakeholder Interactions in an Innovation Ecosystem

From our interviews, we found that stakeholder interactions are mediated through three key commodities: money (M), people (P) and knowledge (K). Described below are outgoing interactions for each of the stakeholders.

*University stakeholders:* These stakeholders are primarily concerned about education and research activities. They interact with corporate stakeholders for educational and research partnerships. These interactions help ensure a pipeline of talent (P) and knowledge (K) from universities to industry. Further, an important theme from the interviews was how universities support entrepreneurs who spin out of the university. The support comes in the form of resources, scientific and business expertise (K), mentors (P), and small amounts of funding through business plan competitions or grants (M). University stakeholders (especially public universities) also interact with government stakeholders to ensure that the university is contributing to the economic prosperity of the region (M). This interaction is relatively indirect because the university's economic contribution is mediated through complex factors such as workforce and education policy, city infrastructure, housing prices etc. Hence this interaction is categorized as a weak interaction.

*Government stakeholders:* Government stakeholders interact with university stakeholders primarily for the purpose of providing funding for research, financial aid, scholarships etc. (M). The government also directly supports entrepreneurs through Small Business Innovation Research (SBIR)/Small Business Technology Transfer (STTR) grants (M).

*Risk Capital:* Risk capital includes venture capital firms, angel investors and seed firms that interact with entrepreneurs to invest money in their enterprise (M). Many risk capital firms also advise entrepreneurs on business strategy based on their own experience in a particular market (K).

*Entrepreneurs:* If an entrepreneur's business is based on intellectual property (IP) or licensing technology developed in a university, then the university gets royalties from the sale of the business's product (M). This interaction is weak because not all entrepreneurs spin-out of universities, so the intensity of these interactions is relatively low. Similarly, if an entrepreneur's business is successful, it often gets acquired by a larger corporation and therefore the corporation often gets rights to the entrepreneur's product (K). However, most entrepreneurs are not successful enough to reach the acquisition/merger stage and so these interactions are relatively infrequent. Lastly, risk capital interacts with entrepreneurs to invest in the enterprise, in return entrepreneurs negotiate financial terms with investors which determine how much return an investor gets on their investment in the business (M).

*Corporate Stakeholders:* Corporate stakeholders (often alumni) interact with universities primarily for the purpose of funding research efforts, donating to the university or sponsoring educational projects. Therefore, the context of these interactions is primarily monetary (M). Corporate stakeholders may also interact with entrepreneurs in an advisory, formal, or informal mentorship role (K). These interactions are classified as weak because often the advisory or mentorship roles are undertaken by an individual at a corporation who may decide to mentor or advise an entrepreneurial enterprise. Corporations do not frequently undertake these activities at an organizational level.

*5.1.3 Information Activities, Tools, and Technologies*

Stakeholder motivations frame the context in which they interact with each other. These interactions result in exchange of people, money, and knowledge between stakeholders and are comprised of various activities such as research, networking, fundraising, etc. These activities can broadly be categorized as information seeking activities, information management activities and activities that synthesize information into actionable goals. Stakeholders used a diversity of tools, technologies, and resources to support various activities.

*Digital Tools*

Digital tools mainly supported information seeking activities and can be divided into the following categories:

*Online Databases/Data Services:* These tools provide information about different companies in a sector or technology area of interest. Examples include Pitchbook, Crunchbase, Statista etc. Pitchbook emerged as the most prominent online data service that participants used to look up startups and their financial information. Participants used Pitchbook to get an idea of how much money was flowing into markets or technologies of interest and who were the key players. It was primarily used by participants who were entrepreneurs, investors, and worked in accelerators or corporations. Other databases included federal databases such as the US Census Data and data by the Department of Labor Statistics, which were mainly used by participants undertaking economic development activities (usually government stakeholders).

*Industry and Market Reports:* A few stakeholders, mainly entrepreneurs, investors and corporate stakeholders used online market reports from firms such as BCC, Gartner, Markets and Markets etc. to obtain the current "lay of the land" of the market they were interested in. These are high-level reports that describe global market trends such as market growth prediction, key barriers and drivers of growth and market size information.

*Academic Articles:* Some stakeholders also used digital databases to access scientific and academic articles to gain knowledge in their area of interest. Entrepreneurs used this for further product development. Government stakeholders used academic articles primarily in economics to obtain metrics of interest (for example GDP, housing prices etc.) or keep up with economic trends. Some university stakeholders such as technology transfer managers use scientific academic articles to compare an innovation disclosure (lab technology) with existing technologies.

*Social Media/Networks:* These digital platforms include LinkedIn, Instagram, technology magazines (such as TechCrunch and Geekwire) and blogs. Stakeholders used these tools primarily to find *people* of interest, market to target audiences and find partnerships. In the context of innovation ecosystems, these tools were primarily used by participants who were entrepreneurs, investors, and worked in accelerators or corporations.

*Organizational:* These digital tools allow stakeholders to manage organizational processes such as keeping track of how many people are served and solicit feedback via surveys. Examples include Salesforce, Google Surveys and Hubspot.

Other tools such as AirTable and Gsuite Office tools were used for collaborative team activities. Digital organizational tools were used by all stakeholders.

*Internal Database:* These are databases maintained internally by various organizations. All stakeholders maintained or used internal databases. The most common digital tool used for this purpose is Microsoft Excel. The information contained within these internal databases differed widely across different stakeholders. These databases usually contain proprietary information, such as finances, human resource tracking, program management metrics etc.

*Non-Digital Resources*

While stakeholders used digital tools for high-level information search, management, and synthesis, all the stakeholders stressed that people, more than digital tools, are the most important source of information, when seeking information in innovation ecosystem. Three categories of people emerged as being as prominent sources of information

*Personal/Professional Network*: These are individuals in the stakeholders' immediate network who they rely on for advice and strategy, partnerships/recruiting or funding.

*Experts:* Most stakeholders rely on experts in their fields (such as scientists, researchers, clinicians etc.) to provide information on the novelty of a technology, inform business strategy, disclose new inventions at the university or solicit advice on partnerships and funding. Experts are also sought after by entrepreneurs, government, and university stakeholders to serve on advisory boards.

*Alumni:* University stakeholders and entrepreneurs who spin out of the university heavily rely on the university's alumni network to find people who can help inform business/program strategy, build, or help solicit partnerships/ and most importantly provide funds and donations. Stakeholders who run university programs especially educational programs for innovation and entrepreneurship seemed to heavily rely on alumni for funding. Alumni are also on advisory committees in universities, government and corporate and startup boards that steer the direction of these organizations.

*Other:* Some stakeholders also mentioned organizations such as accelerators, incubators, consulting organizations or trade associations as being an important source of information, specifically in the context of business and entrepreneurship.

Table 4 summarizes how different tools and resources map to different activities and stakeholders. Note that if a stakeholder category is not mapped to a tool, it does not imply that the tool was not at all used by anyone in that group. The mapped categories imply that these groups actively used the tool enough for them to mention it during the interviews. From the table, it is evident that most of the digital tools support information seeking activities with some information management and synthesis. All stakeholders used some form of online database, internal database, and organizational tools to support information seeking, management and synthesis activities. All stakeholders also leveraged their personal and professional network to gain various types of information.

An important finding of our study is that all stakeholders commented that digital information-seeking tools are good at providing "high-level" information. However, they usually depend on their network and people when making critical decisions. Stakeholders use the general (for example, general market size estimates, relevant companies in a particular field etc.) information obtained from digital tools as a starting point to frame their decision-making activities. However, they ultimately rely on the contextual information (for example, how to price products for a particular market, the credibility of a potential partner etc.), obtained from people to take real action. Figure 3 provides a framework describing how stakeholders use digital and non-digital tools and resources to operate in an innovation ecosystem.

Table 4: A mapping of stakeholder groups and activities to tools and resources used

| Tools & Resources | Activities | Activity Type | Stakeholders |
|---|---|---|---|
| Online Databases | Look up financial information about other companies, find other companies to contact, economic development metrics, relevant statistics | Information seeking | All |
| Industry and Market reports | Get general "high-level" information about market trends and predictions | Information seeking | Entrepreneurs, corporate, risk capital |
| Academic Articles | Get information on emerging technological or economic trends | Information seeking | Entrepreneurs, government, university |
| Social Media and Networks | Outreach (scout for companies to be part of a program), look for key thought leaders, advisors, and mentors to connect with. | Information seeking | Entrepreneurs, risk capital, corporate |
| Organizational Databases | Track leads (entrepreneurs and risk capital), conduct satisfaction/feedback surveys, collaborate across teams | Information seeking, management and synthesis | All |
| Internal Databases | Maintain historical records (financial and people-based), maintain a repository of resources (articles, process documents etc.) | Information management | All |
| Personal/Professional networks | Look for key thought leaders, advisors, and mentors to connect with, getting advice and mentorship | Information seeking | All |
| Experts | Get information on topics that require technical, business, or other expertise | Information seeking | Entrepreneurs, university, government |
| Alumni | Look for key thought leaders, advisors, and mentors to connect with, getting advice and mentorship, scouting for donors and project/program sponsors. | Information seeking | University, entrepreneurs (university spinouts) |

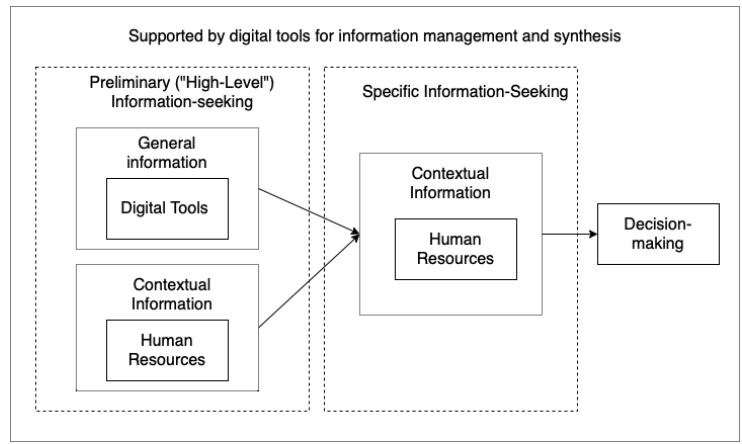

Figure 3: A framework for information-seeking practices of stakeholders in an innovation ecosystem as mediated by digital and non-digital tools and resources

Figure 4 illustrates the relationship between stakeholder motivations, interactions, activities, tools, and resources that enable stakeholder decision-making.

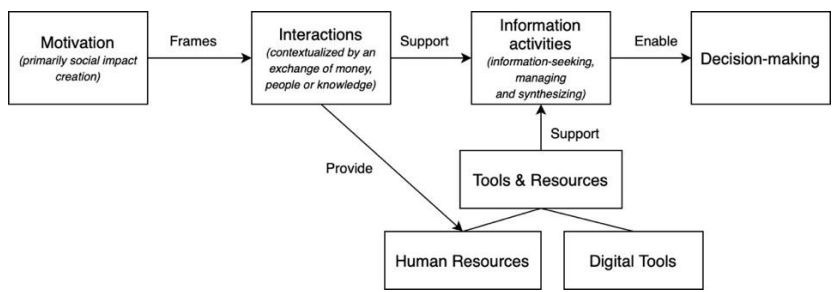

Figure 4: A summary of how stakeholder motivations, interactions, activities, tools, and technologies interact to enable decision-making in innovation ecosystems

### 5.2 Information Needs of Stakeholders

Prior results provided a general lens of stakeholder motivations, information activities and the tools they used. Through Part 2 of the interview, we dove deeper to understand *what type* of information stakeholders are typically interested in. Specifically, we asked participants to "think out loud" as they navigated a data visualization tool we created, which consisted of general quantitative metrics used to describe an innovation ecosystem (See Section 4).

Table 5 summarizes the data types on the tool that most and least interested different stakeholder groups. Note the above mapping are not absolute. Some stakeholders might be using metrics that are generally not of interest to the rest of the group and vice versa. The mappings assume that the data types that stakeholders spent most time on or commented the most on were of most interest and vice versa.

Table 5: Information Needs of Stakeholder Groups in Innovation Ecosystems

| Stakeholder Groups | Data Types of Interest | Data Types Not of Interest |
|---|---|---|
| Entrepreneurs | Federal R&D, Venture Capital Funding by sector and technology, total venture capital funding | Employment, Establishments data SBIR |
| University Stakeholders | Federal R&D, SBIR, Employment, Venture Capital Funding by sector, total venture capital funding (mild interest) | Venture Capital Data (for around half the stakeholders), Establishments |
| Government Stakeholders | Employment Data | Venture Capital Data, Federal R&D Data |
| Risk Capital Stakeholders | Federal R&D Data, SBIR, Venture Capital Funding by sector and technology, total venture capital funding | Employment (for around half the stakeholders), Establishments |
| Corporate Stakeholders | Federal R&D, SBIR (mild interest), Venture Capital Funding by sector and technology, total venture capital funding, Employment (mild interest) | Establishments Data |

#### 5.2.1 Similarities in Stakeholders' Information Needs

***Visualization tool is useful:*** Almost all stakeholders (*N = 11*) said that having a visualization tool that allows them to interact with the given metrics will be useful, especially after further design iterations. Currently, the information seeking activities of all stakeholders is scattered across different digital tools and human resources. Therefore, anything that can reduce the effort of scouring various digital sources while supporting coherent decision-making is welcome by stakeholders. For example, one of the stakeholders exclaimed that

"Pitchbook doesn't capture Federal R&D funding or SBIR funding, so this could help with that".

*Comparative Analysis:* All stakeholders *(N = 13)* liked being able to compare different metrics across different cities. Stakeholders expressed that the comparative analysis could aid them in targeting their activities (marketing, sales, outreach, partnership, investment etc.) in a data-driven manner, as one stakeholder commented:

"If this is a polished product, it'll save a lot of time".

Another stakeholder explained how a tool like this would help in streamlining their activities:

"[Having this type of a tool] definitely helps in having a focused and targeted (sales, marketing, investment) effort and not wasting effort and money in areas that won't be that useful"

*Establishments Data:* Almost all stakeholders *(N = 11)* either did not comment on the benefit of knowing the breakdown of companies in a city by sector or explicitly stated that they did not find this metric to be useful. Only around a third of the stakeholders navigated to this page (*N= 4*). Out of those who did, only two of them verbalized interesting insights they gained from the information presented on this page.

5.2.2 Differences between Stakeholders' Information Needs

*Employment Data:* Most university and government stakeholders were interested in employment trends in the region. University stakeholders were interested in employment trends from an educational perspective, to see where different students go and which industries need more employees. Similarly, government stakeholders were interested in employment from a workforce perspective, to see where the regional workforce could be developed. Entrepreneurs were less interested or skipped over looking at employment trends entirely. Risk capital and corporate stakeholders, either showed no interest in employment trends or showed mild interest with employment being a proxy to determine the state of industries in different regions. Given below is an example by one risk capital stakeholder:

"If I were a scout for a fund or worked for a fund and wanted to increase my deal flow and let's say I am investing in manufacturing. I would use the tool and say okay there a lot of manufacturing workforce in Los Angeles (LA). So, LA probably have some of the facilitating factors to incubate a startup ecosystem around manufacturing."

*Technology Verticals:* In contrast to employment data, most university and government stakeholders were not as interested in how much funding was allocated to different technology verticals (FinTech, Clean Energy etc.). On the other hand, entrepreneurs, risk capital, corporate and accelerator stakeholders found this information to be useful. Entrepreneurs wanted to know this information to see which regions are better funded for the technology vertical that aligns with their enterprise. Corporate stakeholders were interested in this information to see the technology profile of different cities for further investments and partnerships. Risk capital stakeholders were interested in this data as it could provide them with a regional analysis of where venture funds are concentrated for technologies that they are interested in funding. Stakeholders who managed accelerators were interested to see what other streams of technology they could add to their program.

## 5.3 Usability Needs of Stakeholders

All stakeholders were unanimous on aspects of the tool that could be improved. The consensus was that just like most digital tools, our visualization tool provided a "high-level" insight into trends. However, contextual information specific to stakeholder needs was missing. The need for specific contextual information was expressed along four dimensions:

*Granularity:* Nine out of 13 stakeholders explicitly stated that for the tool to be useful to them, they would like to see more granular information, or "details-on-demand" [28] but in context. Examples of granularity included additional data on breaking down the federal R&D funding further by funding agency (NASA, NIH, NSF etc.) or breaking the industry segments further into its constituents. For example, healthcare information could have been further broken down into digital health, pharmaceuticals, drug discovery etc. As one stakeholder put it:

> "In a perfect world, it would break it down to the level of detail where it showed company names that run into some of these categories, because you know that's the type of information that our educational program is looking for to answer the questions: "Do we do we already have them on the list? Do we need to find a contact at xyz company?"

*People/organizations of interest:* Seven of the 13 stakeholders said that this tool would truly be useful to them if they could use it to find relevant people/organizations and their contact information. This finding validates the information activities framework (Figure 2) from part 1, where stakeholders place more importance on finding the right people to get information instead of relying completely on digital data and tools. For example, one university stakeholder commented:

> "For us just knowing where the entrepreneurship centers are you know, give us an idea of who might be good partners to learn from, rather than having to go through websites and searching"

*Industry Specific Information:* Six out of 13 stakeholders expressed the need for the data to be more specific to their industry of interest. For example, one of the stakeholders working in healthcare technologies wanted more information about a region's strengths in different medical fields as determined by the number of clinicians in that field. In this case, the context is provided by the industry of interest. This was aptly summarized by one stakeholder:

> "How valuable is knowing about basic research funding, knowing about SBIR funding, and then knowing about VC funding? Because those are different kinds of activities we're talking about. If we're talking about life sciences, it matters. If you're talking about B2C (business-to-customer) tech tools, they don't really care about the basic research funding as much. They might much more care about the employment pieces. The utility of all that information is sector specific. If you want to extract a more general principle it's businesses where basic science research really matter and where the technology is the differentiator, that's where basic research matters."

*Personalization:* Five out of 13 stakeholders would have liked the tool to be more "personalized" or "customizable". One of the stakeholders suggested using keywords to search for industry sectors and technology verticals, like Pitchbook does. Another common feedback was to be able to compare metrics between two cities of the users choosing. Stakeholders also suggested being able to generate a custom "report" where they could pick and choose metrics of interest and the tool could put it together in a coherent, exportable format. Some stakeholders suggested adding some form of recommendation system to answer questions that they might have. For example:

"Given you probably know my race and gender and that I am a person say in the 30s, can you tell me which of the venture funds fit my profile and what are the best match for me? That's the data I would like to see for this to be really useful."

The bottom-line was that stakeholders wanted to be able to sort, filter, select and export metrics that pertained to their context and therefore customize the tool to their use case.

### 5.4 Summary of Results

To frame stakeholder needs and practices in an innovation ecosystem, we first examined stakeholder motivations, interactions, activities, tools, and resources that framed the rest of the discussion. These motivations informed stakeholder interactions with each other and the digital tools they used. The motivations and interactions also dictated various activities that stakeholders conducted in their jobs. These activities were supported by a diverse set of digital tools and human resources. We found that digital tools and technologies scaffolded stakeholder efforts by providing a preliminary layer to their information seeking activities. Digital tools also supported information management and synthesis activities. Information from digital tools was fed into a "filter" of human resources which provided a contextual lens to the general information collected, allowing stakeholders to make better informed decisions. We then delved into specific information needs of different stakeholders. Table 5 summarizes the specific information needs of different stakeholders. A key finding from this section was the stakeholders wanted more contextual information from the tool such as more granularity to the data, more industry specific information, being able to personalize their user experience and importantly be able to find relevant human resources for further activities.

## 6 DISCUSSION

### 6.1 Challenges

An innovation ecosystem is composed of three fundamental resources: talent (people), knowledge and capital (money) [1]. Creating, finding, retaining, and circulating these three resources in part characterizes how an innovation ecosystem works. and form the basis of key challenges stakeholders face in an innovation ecosystem:

*Talent (People):* As revealed in the results (Section 5), people are the most sought-after resource in an innovation ecosystem. Talent can be accessed in two ways: (1) it can be grown by investing in educational infrastructure and (2) it can be attracted or found from other regions or organizations. This paper focuses on the latter aspect of accessing talent. Stakeholders used a mix of digital tools (LinkedIn, Instagram etc.) to find the *right* people or leverage their personal/professional networks to do so. Accessing the implicit knowledge embedded in people is one of the main challenges stakeholders faced.

*Knowledge:* The types of knowledge stakeholders seemed to need was explicit knowledge and tacit knowledge. Explicit knowledge is knowledge that is easily articulated, codified, and structured [30]. Examples of explicit knowledge that stakeholders used were academic research articles, textbooks, media articles, blogs, and online databases. Tacit knowledge refers to knowledge that is difficult to express and transfer. This is often embedded in people and is the source of contextual information that many stakeholders seek. The difficulty of accessing tacit knowledge has been highlighted previously when discussing challenges in accessing talent. However, stakeholders had also expressed having difficulty accessing explicit knowledge. These were challenges mainly pertaining to equity as many sources of explicit knowledge (academic journal articles, market reports, subscriptions to online databases) are prohibitively expensive.

*Money (Capital):* Money is needed to support the infrastructure that forms an innovation ecosystem. Examples of capital needs different stakeholders expressed include money for growing a business, money for funding educational programs, money for setting up accelerator programs, money for conducting research etc. One of the key challenges for stakeholders is to raise money for whatever activity they are conducting. While stakeholders use digital tools to identify potential leads for raising capital, the actual deals and negotiations are usually conducted through people.

Effective and equitable access to talent, knowledge and capital ensures that *all* stakeholders in an innovation ecosystem are getting what they need to make their best contributions. The challenges highlighted above emphasize existing inefficiencies in an innovation ecosystem that could be addressed through well-designed technological solutions. However, the first step to designing effective tools is to understand stakeholders' existing practices for obtaining knowledge, talent, and money. Specifically, we need to (1) understand why stakeholders rely more on people than on the digital tools they use, and (2) how can we design effective digital tools that enhance and integrate with existing stakeholder practices.

**6.2 The Role of Trust and Experience**

A key finding of our study was that the information obtained from different sources was not given the same level of importance by the stakeholders. Specifically, information obtained from human resources (personal/professional networks, experts, alumni etc.) was deemed to be *more valuable* than information obtained from digital tools. Almost all participants emphasized that no digital tool could compare with the power of personal and professional networks. As one of the stakeholders commented:

> "One of the values of an in-person network is the ability to iterate on your knowledge-seeking via conversation. It's the kind of thing that even email isn't as good at about. When you ask someone a question, they say something that makes you think about something else. It's like doing follow up questions in an interview. You can't plan for every question you will ask because it depends on what the interviewee is going to say. Same with learning something new. When learning something new, you often don't even know what the questions are that you need to as. That's the advantage of a real-time interaction."

Two key reasons emerged as to why information from people was valued over digital tools: *trust* and *experience*.

*Trust:* Stakeholders relied on their personal/professional networks, experts, alumni, or other resources because they *trusted* the information they were obtaining. Stakeholders had often spent years building relationships with individuals in their network. For example, one stakeholder mentioned that he prefers to leverage his network that he had *"painstakingly built over 15 years"* to get the information he needs. In the case of experts, trust emerged from the experts' credibility in their field of expertise. For alumni, trust is brokered by a sense of belonging to the same institution. In the context of innovation and business, the role of close social ties and resultant increase in trustworthiness has been previously established in literature. [31,32].

*Experience:* Stakeholders placed a lot of value on the experience their network had in successfully conducting similar activities in the past. Stakeholders valued experience because it was often not just limited to a technical or business area but also extended to the experience an individual had in navigating the social, political, cultural, and economic aspects of a particular innovation ecosystem. Therefore, the information obtained from such a resource was deemed to be more nuanced and accurate as it was in the right context. As one entrepreneur said:

> "The advisors have decades of experience, so not only are they well-versed in the technical ins and outs, but also the politics between institutions, interpersonal situations etc. Sometimes the advisors might not know the answer themselves but might know someone else in their network who can help and so the advisors can connect us to that resource."

All stakeholders implicitly or explicitly stated that the contextual nature of the knowledge they are seeking tends to come with experience and cannot be easily replicated digitally. Therefore, they rely on other people who have experience in the desirable context to obtain relevant information. Our findings reinforce prior work, which emphasizes the importance of networks of stakeholders with diverse experiences and expertise in innovation ecosystems [33,34,35].

Therefore, the reason stakeholders value information from people more than digital tools is because they trust the information they obtain from people in their network and believe that the experience people have provides better contextual information than any digital tool. So, the question becomes: how do we design tools that complement existing stakeholder practices of leveraging their network built on trust and experience?

**6.3 Design Considerations: How can technology help in enhancing stakeholder decision-making?**

As one stakeholder commented:

"Data and technology cannot capture spontaneity and how people interact "

This comment elicits the fundamental socio-technical gap in computer supported technologies [36], the gap between what social actors need and what technology can provide. In the context of innovation, the goal of technology should not be to replace the spontaneity, uncertainty, and network dynamics inherent in the innovation process. Rather, technology can be used to enable stakeholders to better access knowledge, human resources and capital and better organize their activities within an innovation ecosystem. To that end, using insights from our interviews we synthesized design tasks and considerations that will inform further iterations of our visualization tool and exemplify a technological solution that alleviates some of the challenges stakeholders face. Table 6 provides a summary of these design tasks, considerations, and example applications for each consideration.

***Support Diverse Stakeholder Contexts:*** In Tables 4 and 5 we outlined the information activities and data needs of different stakeholder groups. These data needs and activity patterns highlight the underlying context that different stakeholders operate in pertaining to their innovation ecosystem. Therefore, further work is needed to develop tool designs that accommodate different stakeholder needs and support the rich interactions occurring in innovation ecosystems. How do we create a customizable tool that accounts for differences in stakeholder needs and activities? Can data type and activity categorization act as effective selection parameters for different stakeholders? Do we design a single tool that caters to different stakeholders or a suite of tools that cater to each stakeholder?

***Support Multiple Modes of Analyses****:* We asked stakeholders whether they preferred a comparative analysis across different cities or in-depth information about a single city. Almost all stakeholders expressed that they would like to access both modes of analyses depending on their information and task needs. Furthermore, stakeholders also expressed the need to be able to view information at higher or lower geographical units (such as states, counties, etc.). What emerged is that most stakeholders combine top-down and bottom-up approaches in their information activities, also known as a middle-out approach. Middle-out design has some precedence in human-computer interaction (HCI). Fredericks et al. proposed middle-out design framework for collaborative community engagement in urban HCI [37]. Middle-out approaches have also been used to develop tools for visual graph analytics [38] and assess models of legal governance for data protection and AI [39]. Future work needs to consider middle-out models of design and answer the question, "how to we develop the visualization tool such that it *seamlessly* transitions between different modes of analyses.

Table 6: Design Tasks and Considerations for Future Tool Design

| Tasks | Considerations | Example Applications |
| --- | --- | --- |
| Support diverse stakeholder contexts | Allow stakeholders to pick metrics relevant to their context and remove other "chart junk" | All stakeholders voiced a need for personalization, in which they wished they could only see data useful in their context, whether it be industry-specific information or stakeholder role specific information. |
| Support multiple modes of analyses | Allow stakeholders to compare relevant statistics between different geographies | A university stakeholder looking to compare the number of spinouts from their university to that of universities in other regions. |
| | Allow stakeholders to dive deeper into one region/city | A risk capital stakeholder who only invests in a specific city might want to know the promising startup, research activity and venture funding landscape of that city. |
| | Make it easy for stakeholders to look at statewide, county-wide, region-wide and city-wide metrics | Economic developments stakeholders are often interested in county-level data to assess the impact of innovation policies |
| Provide information on people and organizations | Clearly highlight who are the key players (people and organizations) in a given context. | All stakeholders expressed a need for knowing who to reach out to, based on the information presented to them. Therefore, they wanted to have some idea of who the key players were in a region and in their context. |
| Drive action-oriented decision-making | Automatically synthesize insights from visualizations of interest | University stakeholders were interested in "fast-facts" such as which institutions are ranked higher than them in a certain area. Therefore, some voiced a need to be able to see this information quickly instead of navigating through the visualization to infer the information. |
| | Make basic recommendations for next steps | All stakeholders wanted to answer the question: "what do I do with this information?" They wanted to get tailored results to what they should do next based on their assumptions |

*Provide Information on People and Organizations:* A key insights from our interviews is that stakeholders value information from their network of people and organizations more than information obtained from digital tools. Therefore, any visualization tool that seeks to support stakeholders in an innovation ecosystem should address the fundamental task of networking and finding key players in an ecosystem. Existing tools such as Pitchbook and Crunchbase provide this information by allowing users to see names and contact information of companies, founders, risk capital organization etc. However, this capability only caters to entrepreneurs, risk capital and corporate stakeholders. In our interviews, we discovered that all stakeholder groups want insight on who are the key leaders in their innovation ecosystem including researchers, scientists, leaders in public institutions etc. An avenue of providing this information can be through visualizing networks of people and organizations. Network visualization is a prominent area of research [40], with applications in business [19], bibliometric analysis [41] to visualize how different fields of research are connected etc. Future iterations of the tool should consider leveraging different visualization techniques such as network visualization to inform how stakeholders can better access the relevant human resources for their information needs.

*Drive Action-Oriented Decision-Making:* Almost all stakeholders stated that in its current form, the visualization tool is good for "reporting" statistics and metrics but does not aid directly in decision-making. For example, some stakeholders alluded to having recommender systems that can answer user queries or generate automated reports for information relevant to their context. Therefore, with future work, we hope to answer the question: "how do we make the leap from a reporting tool to a tool that enables action?"

The above design tasks and considerations do not seek to *replace* non-digital sources of information (namely, people) in an innovation ecosystem. Rather, the design considerations focus on developing an intuitive, *user-centered* tool that makes digital information more accessible and useful to stakeholders. This is so that stakeholders make *context-aware* decisions on the knowledge and money needed to fuel further decision-making and know who to approach for it.

### 6.4 Critical Reflections: The Equity Problem

A theme that was implicit in our interviews was that of equity. Throughout different conversations, we paid special attention to the question of "who is able to innovate and participate in innovation ecosystems?". It emerged that innovation ecosystems are inherently inequitable. These inequities were systemic and permeated into stakeholder interactions with both digital tools and human resources.

A key source of inequity was the social and generational capital required to be an entrepreneur or risk capital stakeholder. One entrepreneur commented how she was advised to raise an initial "friends and family" round of funding, where she sought funding through personal network and family ties. However, she commented that she did not have friends and family who were able to fund her enterprise, which potentially set her back. Similarly, social, and generational capital also determines who an individual is connected to [42], which plays an important role in who an individual can reach out to for their information needs. Those who do not have adequate social and generational capital to support their innovative activities must do an immense amount of work upfront to build that kind of capital.

A related source of inequity is systemic underrepresentation of minorities in related disciplines. In this paper, we are considering "high-technology" and STEM industries. The fact the minorities are underrepresented in these fields is no secret [43]. Therefore, who goes through the STEM pipeline as a student, scientist, employee, entrepreneur etc. has downstream effects on who is able to participate in an innovation ecosystem.

Finally, inequity in the context of digital tools exist in the cost associated with these tools. For example, an annual license for Pitchbook costs $25,000 [44]. Similarly, market and industry report costs are in the range of $1500-$8000 each [45]. Additionally, different academic publishing venues also impose annual subscription costs to access articles [46,47]. Therefore, the pricing of these tools indicates that they are not built to support low-resource stakeholders who do not have the means or organizational ties (for example, being at a university) to purchase these digital services.

Therefore, further tool design should consider the above angles of inequities built into the system and attempt to leverage technology to level the playing field for participating in an innovation ecosystem. This involves answering the question: how do we design *accessible* digital tools that facilitate equitable access of knowledge, people, and capital to stakeholder from diverse backgrounds?

## 7 LIMITATIONS

There are three key limitations of this study. First, all the participants of the study belonged to the Seattle innovation ecosystem. Given the large variability between innovation ecosystems, this is advantageous in endowing depth and a fixed regional context to the analysis. However, it also means that our findings may only be specific to one region. In our future work, we will expand our study to other regions sampled from the metropolitan areas which were considered in the digital dashboard. Second some participants had overlapping roles, which affected how they used and interpreted the digital dashboard and made it difficult to disentangle the perspective they spoke from. Lastly, our study had a limited number of participants and diversity in each stakeholder group. An innovation ecosystem consists of many stakeholders with intersecting roles. It is incredibly challenging to exhaustively consider all of them. That being said, in our future work we will consider (a) adding more participants to existing groups for a greater depth of perspective and (b) adding university researchers, students and industry trade associations as participants as they are also contributors to an innovation ecosystem

## 8 CONCLUSION

In this study, we investigated stakeholder motivations, interactions, activities, tools, and resources used in an innovation ecosystem. Through a three-part interview study with 13 participants from six stakeholder groups and digital dashboard tool, we studied (1) how stakeholder groups seek to contribute to and interact within innovation ecosystems, (2) what are their information-seeking practices and needs to support their contributions and interactions, and (3) what are the design considerations for developing effective digital tools that support those needs and practices. We found that stakeholders' primary motivation to participate was the social impact of their contributions which framed stakeholder interactions. These interactions were based on the exchange of people, money, and knowledge between stakeholders. Consequently, most information needs were focused on engaging with the right people to obtain money and knowledge in the right context. Stakeholders used a variety of digital tools and non-digital resources to seek information. A key finding of our study was that stakeholders used digital tools to seek "high-level" information to scaffold initial decision-making efforts but ultimately relied on contextual information provided by human networks to enact final decisions. This is because stakeholders had established trust with other people in their network and believed that the information they need came with experience in the right context, which people have, and digital tools lack. Based on our findings, we proposed 5 key design tasks and 7 considerations for designing future analytic tools to support stakeholders in innovation ecosystems. In future work, we hope to iterate on the current dashboard tool based on the design considerations proposed in this study and evaluate the next iteration by continual engagement with multiple stakeholder groups.

## REFERENCES

[1] National Research Council. 2012. Rising to the Challenge. National Academies Press.
[2] Nick Henry, Tim Angus, and Mark Jenkins. 2021. Motorsport Valley revisited: Cluster evolution, strategic cluster coupling and resilience. European Urban and Regional Studies 28, 4 (2021), 466-486. DOI:https://doi.org/10.1177/09697764211016039
[3] Amnon Frenkel, Shlomo Maital, Eran Leck, Daphne Getz, and Vered Segal. 2011. Israel's Innovation Ecosystem. Israel: Samuel Neaman Instute 30 (2011).
[4] AnnaLee Saxenian. 1996. Regional Advantage. Harvard University Press.
[5] Margaret O'Mara. 2020. The Code. Penguin.
[6] Henry Etzkowitz and Loet Leydesdorff. 2000. The dynamics of innovation: from National Systems and "Mode 2" to a Triple Helix of university–industry–government relations. Research Policy 29, 2 (2000), 109-123. DOI:https://doi.org/10.1016/s0048-7333(99)00055-4
[7] Christopher Freeman and Luc Soete. 2009. Developing science, technology and innovation indicators: What we can learn from the past. Research Policy 38, 4 (2009), 583-589. DOI:https://doi.org/10.1016/j.respol.2009.01.018
[8] Wesley M. Cohen, Richard R. Nelson, and John P. Walsh. 2002. Links and Impacts: The Influence of Public Research on Industrial R&D. Management Science 48, 1 (2002), 1-23. DOI:https://doi.org/10.1287/mnsc.48.1.1.14273
[9] Dominique Foray and Francesco Lissoni. 2010. University Research and Public–Private Interaction. Handbook of The Economics of Innovation, Vol. 1 (2010), 275-314. DOI:https://doi.org/10.1016/s0169-7218(10)01006-3
[10] David C. Mowery and Bhaven N. Sampat. 2006. Universities in National Innovation Systems. Oxford Handbooks Online (2006). DOI:https://doi.org/10.1093/oxfordhb/9780199286805.003.0008
[11] Phil Budden and Fiona Murray. 2019. MIT's stakeholder framework for building & accelerating innovation ecosystems. MIT Lab for Innovation Science and Policy. Retrieved September 13, 2022 from https://innovation.mit.edu/assets/MIT-Stakeholder-Framework_Innovation-Ecosystems.pdf
[12] OECD and Eurostat. 2018. The Measurement of Scientific, Technological and Innovation Activities Oslo Manual 2018 Guidelines for Collecting, Reporting and Using Data on Innovation, 4th Edition. OECD Publishing.
[13] Ove Granstrand and Marcus Holgersson. 2020. Innovation ecosystems: A conceptual review and a new definition. Technovation 90, (2020), 102098. DOI:https://doi.org/10.1016/j.technovation.2019.102098
[14] Jason Owen-Smith and Walter W. Powell. 2006. Accounting for Emergence and Novelty in Boston and Bay Area Biotechnology*. Cluster

Genesis (2006), 61-84. DOI:https://doi.org/10.1093/acprof:oso/9780199207183.003.0004

[15] Brad Feld. 2012. Startup Communities. John Wiley & Sons.

[16] Aleksandra Kuzior, Iryna Pidorycheva, Viacheslav Liashenko, Hanna Shevtsova, and Nataliia Shvets. 2022. Assessment of National Innovation Ecosystems of the EU Countries and Ukraine in the Interests of Their Sustainable Development. Sustainability 14, 14 (2022), 8487. DOI:https://doi.org/10.3390/su14148487

[17] Per L. Bylund. 2016. Mark Zachary Taylor, The politics of innovation: why some countries are better than others at science and technology. Public Choice 170, 3 (2016), 327-329. DOI:https://doi.org/10.1007/s11127-016-0397-5

[18] Mariana Mazzucato. 2015. The Entrepreneurial State. Anthem Press.

[19] Rahul C. Basole, Arjun Srinivasan, Hyunwoo Park, and Shiv Patel. 2018. ecoxight: Discovery, Exploration, and Analysis of Business Ecosystems Using Interactive Visualization. ACM Transactions on Management Information Systems 9, 2 (2018), 1-26. DOI:https://doi.org/10.1145/3185047

[20] Rahul C. Basole, Trustin Clear, Mengdie Hu, Harshit Mehrotra, and John Stasko. 2013. Understanding interfirm relationships in business ecosystems with interactive visualization. IEEE Transactions on Visualization and Computer Graphics 19, 12 (2013), 2526–2535. DOI:http://dx.doi.org/10.1109/tvcg.2013.209

[21] Sebastien Heymann and Benedicte Le Grand. 2013. Visual Analysis of Complex Networks for Business Intelligence with Gephi. 2013 17th International Conference on Information Visualisation (2013). DOI:https://doi.org/10.1109/iv.2013.39

[22] National Science Foundation - State Indicators. Retrieved September 13, 2022 from https://ncses.nsf.gov/indicators/states/

[23] Kauffman Indicators of Entrepreneurship – The Kauffman Indicators . Retrieved September 13, 2022 from https://indicators.kauffman.org/

[24] Innovation Intelligence: StatsAmerica. Retrieved September 13, 2022 from https://www.statsamerica.org/innovation/

[25] Duncan A. Smith. 2016. Online interactive thematic mapping: Applications and techniques for socio-economic research. Computers, Environment and Urban Systems 57, (2016), 106-117.

[26] Innovation Intelligence: StatsAmerica. Retrieved September 13, 2022 from https://www.statsamerica.org/innovation/

[27] Sergi Fàbregues and Marie-Hélène Paré. 2007. Charmaz, Kathy C. (2006). Constructing Grounded Theory: A Practical Guide Through Qualitative Analysis. Papers. Revista de Sociologia 86, (2007), 284. DOI:https://doi.org/10.5565/rev/papers/v86n0.825

[28] Irene Ramos-Vielba, Manuel Fernández-Esquinas, and Elena Espinosa-de-los-Monteros. 2009. Measuring university–industry collaboration in a regional innovation system. Scientometrics 84, 3 (2009), 649-667. DOI:https://doi.org/10.1007/s11192-009-0113-z

[29] Ben Shneiderman. 2003. The Eyes Have It: A Task by Data Type Taxonomy for Information Visualizations. The Craft of Information Visualization (2003), 364-371. DOI:https://doi.org/10.1016/b978-155860915-0/50046-9

[30] Edward R. Tufte. 1983. The Visual Display of Quantitative Information.

[31] Wenpin Tsai and Sumantra Ghoshal. 1998. Social Capital and Value Creation: The Role of Intrafirm Networks. Academy of Management Journal 41, 4 (1998), 464-476. DOI:https://doi.org/10.5465/257085

[32] Brian Uzzi. 1996. The Sources and Consequences of Embeddedness for the Economic Performance of Organizations: The Network Effect. American Sociological Review 61, 4 (1996), 674. DOI:https://doi.org/10.2307/2096399

[33] How to Create an Innovation Ecosystem. Retrieved September 13, 2022 from https://hbr.org/2012/12/how-to-create-an-innovation-ec

[34] Walter W. Powell and Stine Grodal. 2006. Networks of Innovators. Oxford Handbooks Online (2006). DOI:https://doi.org/10.1093/oxfordhb/9780199286805.003.0003

[35] Olav Sorenson. 2018. Innovation Policy in a Networked World. Innovation Policy and the Economy 18, (2018), 53-77. DOI:https://doi.org/10.1086/694407

[36] Harry Collins. 2010. Tacit and Explicit Knowledge. University of Chicago Press.

[37] Mark S. Ackerman. 2000. The Intellectual Challenge of CSCW: The gap between social requirements and technical feasibility. Human–Computer Interaction 15, 2-3 (2000), 179–203. DOI:http://dx.doi.org/10.1207/s15327051hci1523_5

[38] Joel Fredericks, Glenda Amayo Caldwell, Martin Tomitsch. 2016. Middle-out Design: Collaborative Community Engagement in Urban HCI. Proceedings of the 28th Australian Conference on Computer-Human Interaction – OzCHI. DOI:10.1145/3010915.3010997

[39] Pak Chung Wong, Patrick Mackey, Kristin A. Cook, Randall M. Rohrer, Harlan Foote, and Mark A. Whiting. 2009. A multi-level middle-out cross-zooming approach for large graph analytics. 2009 IEEE Symposium on Visual Analytics Science and Technology (2009). DOI:https://doi.org/10.1109/vast.2009.5333880

[40] Ugo Pagallo, Pompeu Casanovas, and Robert Madelin. 2019. The middle-out approach: assessing models of legal governance in data protection, artificial intelligence, and the Web of Data. The Theory and Practice of Legislation 7, 1 (2019), 1-25. DOI:https://doi.org/10.1080/20508840.2019.1664543

[41] I. Herman, G. Melancon, and M.S. Marshall. 2000. Graph visualization and navigation in information visualization: A survey. IEEE Transactions on Visualization and Computer Graphics 6, 1 (2000), 24-43. DOI:https://doi.org/10.1109/2945.841119

[42] Nees Jan Van Eck and Ludo Waltman. 2014. Visualizing Bibliometric Networks. Measuring Scholarly Impact (2014), 285-320. DOI:https://doi.org/10.1007/978-3-319-10377-8_13

[43] Tonia Warnecke. 2013. Entrepreneurship and Gender: An Institutional Perspective. Journal of Economic Issues 47, 2 (2013), 455-464. DOI:https://doi.org/10.2753/jei0021-3624470219

[44] STEM Jobs See Uneven Progress in Increasing Gender, Racial and . Retrieved September 13, 2022 from https://www.pewresearch.org/science/2021/04/01/stem-jobs-see-uneven-progress-in-increasing-gender-racial-and-ethnic-diversity/

[45] PitchBook Pricing 2022. Retrieved September 13, 2022 from https://www.trustradius.com/products/pitchbook/pricing

[46] 4 Tips for Determining Your Market Research Budget. Retrieved September 13, 2022 from https://blog.marketresearch.com/4-tips-for-determining-your-market-research-budget

[47] IEEE Membership and Society Membership Dues - IEEE. Retrieved September 13, 2022 from https://www.ieee.org/membership/join/dues.html

[48] Non-Member Subscription Prices for Individuals. Retrieved September 13, 2022 from https://www.acm.org/publications/alacarte/journalala